\documentclass{jpp}

\usepackage{amsmath}
\usepackage{bm}
\usepackage[T1]{fontenc}
\usepackage{graphicx}
\usepackage{hyperref}
\usepackage[utf8]{inputenc}
\usepackage{textgreek}

\newcommand{\iotabar}{\mbox{$\,\iota\!\!$-}}

\title{Magnetic Fields with General Omnigenity}

\author{Daniel W. Dudt\aff{1}, Alan G. Goodman\aff{2}, Rory Conlin\aff{1}, Dario Panici\aff{1}, Egemen Kolemen\aff{1,3} \corresp{\email{ekolemen@princeton.edu}}}

\affiliation{
\aff{1}Princeton University, Princeton, New Jersey 08544, USA
\aff{2}Max-Planck-Institut für Plasmaphysik, D-17491 Greifswald, Germany
\aff{3}Princeton Plasma Physics Laboratory, Princeton, New Jersey 08543, USA
}

\begin{document}

\maketitle

\begin{abstract}
Omnigenity is a desirable property of toroidal magnetic fields that ensures confinement of trapped particles. 
Confining charged particles is a basic requirement for any fusion power plant design, but it can be difficult to satisfy with the non-axisymmetric magnetic fields used by the stellarator approach. 
Every ideal magnetohydrodynamic equilibrium previously found to approximate omnigenity has been either axisymmetric, quasi-symmetric, or has poloidally closed contours of magnetic field strength $B$. 
However, general omnigenous equilibria are a much larger design space than these subsets. 
A new model is presented and employed in the DESC stellarator optimization suite to represent and discover the full parameter space of omnigenous equilibria. 
Although exact omnigenity aside from quasi-symmetry is impossible, these results reveal that excellent particle confinement can be achieved in practice. 
Examples far from quasi-symmetry with poloidally, helically, and toroidally closed $B$ contours are attained with DESC and shown to have low neoclassical collisional transport and fast particle losses. 
\end{abstract}

\section{Introduction}
\label{sec:intro}

Controlled nuclear fusion offers the possibility of reliable power generation without greenhouse gas emissions to satiate the growing global energy demand. 
The magnetic confinement approach to fusion involves containing a high temperature plasma for long periods of time, and requires a magnetic geometry than can confine these energetic charged particles. 
The stellarator concept \citep{Spitzer1958} makes an attractive fusion reactor: its reliance on external magnetic fields rather than driving plasma current to confine the plasma enables it to operate at steady state and makes it less susceptible to plasma disruptions \citep{Helander2012}. 
Yet historically, stellarators were not pursued due to their relatively high levels of neoclassical transport. 
Unlike its axisymmetric counterpart the tokamak, stellarators are not guaranteed to confine trapped particles in a collisionless plasma. 
Careful attention must be paid to design a stellarator with comparably good particle confinement, but this is becoming increasingly achievable with modern optimization tools. 

\begin{figure}
\centering
\includegraphics[width=0.5\textwidth]{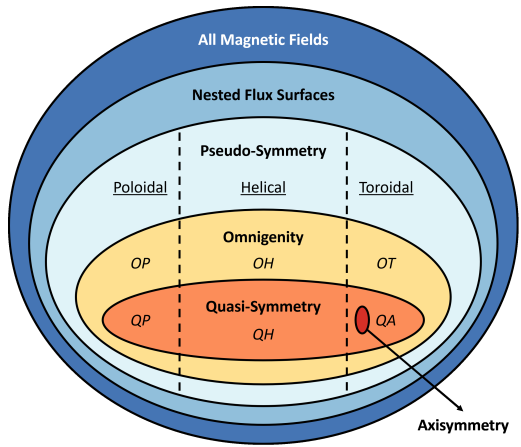}
\caption{Classification of magnetic fields. 
The decreasing area of the inner subsets represents the smaller dimensionality of these subspaces.
Note that OP refers to the same set of magnetic fields that are sometimes labeled QI.}
\label{fig:diagram}
\end{figure}

Trapped particles cannot be avoided altogether in a toroidal geometry. 
If the radial drifts of trapped particles vanish everywhere it is called an \textit{isodynamic} magnetic field, but this is similarly unrealistic \citep{Palumbo1968,Helander2014}. 
The next best theoretical option is \textit{omnigenity}, the class of magnetic fields in which the bounce-averaged radial drifts of trapped particles vanish \citep{Hall1975,Cary1997b}. 
\textit{Pseudo-symmetry} is a broader category of fields in which the contours of constant magnetic field strength $B=|\mathbf{B}|$ on a flux surface are closed curves \citep{Isaev1999}. 
Omnigenity implies pseudo-symmetry and also that the $B_{\mathrm{max}}$ contour is straight in Boozer coordinates (and other straight field line coordinate systems with a Jacobian that only depends on $B$ and the flux surface label) \citep{Helander2014}. 
There are three classes of omnigenous magnetic fields: the $B$ contours can close either poloidally, helically, or toroidally. 
These cases will be referred to as \textit{omnigenity with poloidal contours} (OP), \textit{omnigenity with helical contours} (OH), and \textit{omnigenity with toroidal contours} (OT). 
[The term \textit{quasi-isodynamic} (QI) has often been used in previous literature to mean omnigenity with poloidally closed $B$ contours. 
This terminology has been inconsistent, however, and is easily conflated with more general omnigenity; QI is avoided in favor of OP in this article for the sake of clarity.] 
\textit{Quasi-symmetry} (QS) is a special subset of omnigenity where all contours of $B$ are straight in these coordinates, not only the $B_{\mathrm{max}}$ contour \citep{Nuhrenberg1988,Rodriguez2020}. 
QS magnetic fields are further subdivided by their helicity: \textit{quasi-poloidal symmetry} (QP), \textit{quasi-helical symmetry} (QH), and \textit{quasi-axisymmetry} (QA) are the quasi-symmetric subspaces of OP, OH, and OT, respectfully. 
The Venn diagram in figure \ref{fig:diagram} visualizes the relationship between these different classes of magnetic fields. 

Since omnigenity is a less restrictive requirement than quasi-symmetry, it provides a larger design space with equivalent neoclassical confinement. 
This extra flexibility could help satisfy additional physics and engineering constraints on a stellarator reactor, such as reduced turbulent transport and simplified coil geometry. 
Unfortunately, \cite{Cary1997a,Cary1997b} proved that the only omnigenous fields that are analytic are those that are quasi-symmetric. 
Moreover, \cite{Garren1991} suggested from a near-axis expansion that exact quasi-symmetry cannot be satisfied beyond a single flux surface. 
This is not a significant restriction in practice, however: QA and QH solutions have been found with good QS throughout a volume \citep{Landreman2022a,Zarnstorff2001,Bader2020,Nuhrenberg1988,Henneberg2019}, and many solutions have been discovered that approach OP while being far from QS \citep{Jorge2022,Mata2022,Goodman2022,Sanchez2023}. 
The performance of QH and OP have also been experimentally verified by the HSX and W7-X stellarators, respectively. 
These devices were optimized to have approximately quasi-symmetric or omnigenous designs, and both experiments have demonstrated reduced neoclassical transport \citep{Canik2007,Beidler2021}. 

These recent results have been restricted to QS or OP magnetic fields, which excludes most of the full omnigenity solution space. 
\cite{Plunk2019} explain that OT and OH magnetic fields (that are not QS) cannot be achieved with a first order expansion in the distance from the magnetic axis, but admit that these solutions could exist more generally. 
These classes of omnigenity have previously been neglected due to the lack of an efficient method to parameterize and explore their optimization space. 
Accessing this larger design space would unlock new possibilities for stellarators, and is worthwhile to consider despite its increased computational complexity. 
This article presents an approach to achieve this goal, which is made possible through the DESC stellarator optimization code suite \citep{DESC,Dudt2020,Panici2023,Conlin2023,Dudt2023}. 
It is similar to the original idea of \cite{Cary1997a}, in the sense that a perfectly omnigenous magnetic field is constructed to have the same form on every field line in terms of some new coordinate. 
This omnigenous field is generalized to any helicity, as was also done by \cite{Landreman2012}. 
The difference between the actual equilibrium magnetic field and the constructed target field is then penalized, like the technique devised by \cite{Goodman2022}. 

In contrast to previous approaches, this article provides a parameterization of the full phase space of all omnigenous magnetic fields, and is not restricted to a near-axis expansion formulation or to stellarator symmetry. 
Section \ref{sec:method} gives a derivation of the omnigenity model and explains how it is used as an optimization target. 
Numerical examples of equilibria obtained from this method are provided in section \ref{sec:examples} for all six classes of omnigenity: OP, QP, OH, QH, OT, and QA. 
This is the first demonstration of good neoclassical confinement throughout a volume for realistic equilibria with arbitrary omnigenous magnetic fields. 
The implications of these results and directions for future work are discussed in section \ref{sec:discussion}. 

\section{Method}
\label{sec:method}

\begin{figure}
\centering
\includegraphics[width=\textwidth]{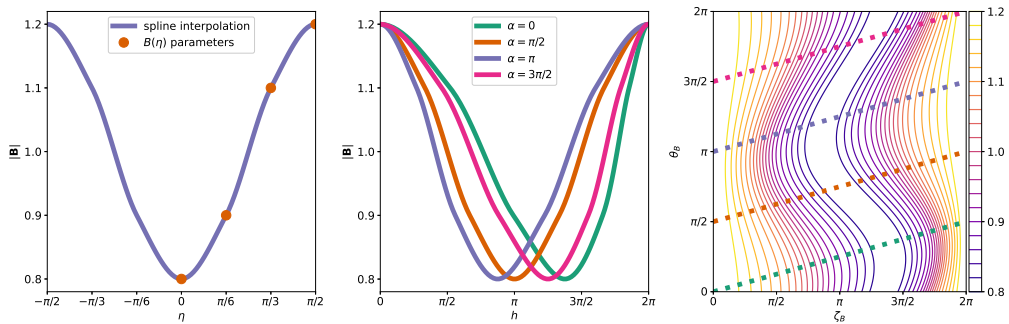}
\caption{Example of an exact omnigenous magnetic field, for an OP case with helicity $M=0,~N=1$ and the parameters $x_{0,1,-1}=-\frac{\pi}{8},~x_{0,1,0}=\frac{\pi}{8},~x_{0,1,+1}=\frac{\pi}{4}$. 
Left: The well shape $B(\eta)$ given by a monotonic spline. 
Center: The magnetic well along different field lines, which all have the same bounce distances $\Delta h$. 
Right: The $B$ contours plotted in Boozer coordinates with a rotational transform of $\iotabar=\frac{1}{4}$.}
\label{fig:model}
\end{figure}

Let the ``helicity'' of the omnigenous magnetic field be defined by the pair of integers $M$ and $N$ such that each $B$ contour closes on itself after traversing the torus $M$ times toroidally
and $N$ times poloidally. 
In addition to having closed contours of magnetic field strength, the ``bounce distance'' $\delta = \sqrt{\Delta\theta_B^2 + \Delta\zeta_B^2}$ in Boozer coordinates $(\theta_B,\zeta_B)$ along a field line between consecutive points with equal magnetic field strength must be independent of the field line label 
\begin{equation}
\label{eq:alpha}
\alpha = \frac{\theta_B - \iotabar \zeta_B}{N - \iotabar M}
\end{equation}
where $\iotabar$ is the rotational transform. 
This normalization has the useful property that $\alpha$ increases by $2\pi$ as a constant-$B$ curve is followed until it closes on itself \citep{Landreman2012}. 

\subsection{Magnetic Well}

First, the ``magnetic well shape'' is parameterized. 
Assuming the magnetic field strength has the same maximum and minimum values along each field line, it can be written in the form 
\begin{equation}
\label{eq:well}
B=B(\rho,\eta)
\end{equation}
where $\rho$ is the flux surface label and contours of constant magnetic field strength are now contours of the new coordinate $\eta$ \citep{Cary1997b}. 
The domain is chosen to be $-\frac{\pi}{2} \le \eta \le \frac{\pi}{2}$ such that $B(0) = B_{\mathrm{min}}$ and $B(\pm\frac{\pi}{2}) = B_{\mathrm{max}}$. 
$B(\eta)$ should be monotonically increasing from $B_{\mathrm{min}}$ to $B_{\mathrm{max}}$ and satisfy the boundary conditions
\begin{equation}
\frac{\partial B}{\partial\eta}\bigg|_{\eta=0} = \frac{\partial B}{\partial\eta}\bigg|_{\eta=\pm\frac{\pi}{2}} = 0.
\end{equation}
$B$ must also be an even function of $\eta$, which will be justified later. 
A magnetic well shape that satisfies these criteria is represented by the parameters
\begin{equation}
\label{eq:B_ij}
B_{ij} = B(\rho_i,\eta_j)
\end{equation}
where the knot locations $\eta_j$ are linearly spaced in the domain $[0, \frac{\pi}{2}]$ and the parameters are sorted to ensure monotonicity. 
$B$ at other locations on the flux surface $\rho_i$ are evaluated by monotonic spline interpolation at $\eta=|\eta|$ to ensure it is an even function. 
The $C^1$ continuity of these monotonic splines is not a limitation since the omnigenous magnetic field is generally not analytic anyways. 
Omnigenity with multiple unique local minima and maxima of $B$ on each flux surface can exist, as shown by \cite{Parra2015}, and this work could be extended to accommodate those cases. 

\subsection{Coordinate Transformation}

Now all that remains is to parameterize the transformation between the Boozer coordinates $(\theta_B,\zeta_B)$ and the computational coordinates $(\eta,\alpha)$. 
This mapping will be defined through the transformation
\begin{equation}
\label{eq:h}
h : (\theta_B,\zeta_B) \leftrightarrow (\rho,\eta,\alpha).
\end{equation}
In Boozer coordinates, the function $h$ is related to the helicity of the omnigenous field: 
\begin{equation}
\label{eq:h_booz}
h(\theta_B,\zeta_B) =
\begin{cases}
N \zeta_B &\text{for }M=0 \\
-\theta_B + \frac{N}{M} \zeta_B &\text{for }M\ne0.
\end{cases}
\end{equation}
[Note that the Boozer angles are an implicit function of $\rho$, since the transformation to Boozer coordinates is unique to each flux surface.] 
Contours of constant $h$ are parallel to the $B_{\mathrm{max}}$ contour (which must be straight in Boozer coordinates), and in the limit of quasi-symmetry $B = B(h)$. 
In the computational coordinates, the function is parameterized as 
\begin{equation}
\label{eq:h_comp}
h(\rho,\eta,\alpha) = 2\eta + \pi + \sum_{l=0}^{L_{\rho}} \sum_{m=0}^{M_{\eta}} \sum_{n=-N_{\alpha}}^{N_{\alpha}} x_{lmn} T_{l}(2\rho-1) F_{m}(\eta) F_{n N_{\mathrm{FP}}}(\alpha)
\end{equation}
where $N_{\mathrm{FP}}$ is the number of field periods, $T_{k}(y)$ are Chebyshev polynomials of the first kind defined as $T_{k}(\cos y) = \cos(k y)$, and $F_{k}(y)$ are Fourier series defined as 
\begin{equation}
F_{k}(y) =
\begin{cases}
\cos(|k| y) &\text{for }k\ge0 \\
\sin(|k| y) &\text{for }k<0.
\end{cases}
\end{equation}
Chebyshev polynomials are chosen as the radial basis functions to allow for an independent well shape on each flux surface. 
[Any set of orthogonal polynomials would suffice. 
Zernike polynomials are not used since they would impose additional unphysical boundary conditions near the magnetic axis.] 
The coefficients $x_{lmn}$ in Eq. \ref{eq:h_comp} parameterize the variation of the well shape on different field lines, and $N_{\alpha}=0$ represents the subspace of quasi-symmetry. 
The boundary condition $h(\rho,\eta=-\frac{\pi}{2},\alpha) = 0$ is required to ensure $B_{\mathrm{max}}$ is a straight contour, which is chosen to be located at $h=0$ by construction. 
This boundary condition is enforced by a linear constraint on the even Fourier modes: 
\begin{equation}
\label{eq:bc}
\sum_{m=0,2,4,\ldots}^{M_{\eta}} (-1)^\frac{m}{2} x_{lmn} = 0.
\end{equation}
This constraint also satisfies the periodicity requirement that $h(\eta=\frac{\pi}{2})=2\pi$ is another $B_{\mathrm{max}}$ contour. 

\subsection{Optimization Objective Function}

The optimization function penalizes the difference between the equilibrium magnetic field $B_{\mathrm{eq}}$ and the constructed omnigenous field $B$, similar to the approach used by \cite{Goodman2022}. 
The errors are evaluated at a series of collocation points in the computational coordinates as: 
\begin{equation}
\label{eq:f_om}
f_{\mathrm{om}}(\rho_i,\eta_i,\alpha_i) = \big[ B_{\mathrm{eq}}(\rho_i,\eta_i,\alpha_i) - B(\rho_i,\eta_i) \big] w(\eta_i)
\end{equation}
where the weighting function $w(\eta_i) = \frac{3}{2} + \frac{1}{2} \cos(2\eta_i)$ is used to weight the minimum of the magnetic well relative to the maximum. 
The omnigenous target field $B$ can be readily evaluated at these collocation points from the magnetic well shape parameters $B_{ij}$ presented in Eq. \ref{eq:B_ij}. 
The equilibrium field $B_{\mathrm{eq}}$, however, must be evaluated at the Boozer coordinates that correspond to the desired computational coordinates, using the mapping defined by Eq. \ref{eq:h}. 

Simultaneously solving Eqs. \ref{eq:alpha} and \ref{eq:h_booz} gives the system of equations 
\begin{subequations}
\begin{align}
\begin{bmatrix}
\alpha \\
h(\rho,\alpha,\eta)
\end{bmatrix}
= \begin{bmatrix}
\frac{1}{N} & -\frac{\iotabar}{N} \\
0 & N
\end{bmatrix}
\begin{bmatrix}
\theta_B \\
\zeta_B
\end{bmatrix}
~&\text{for }M=0, N\neq0 \\
\begin{bmatrix}
\alpha \\
h(\rho,\alpha,\eta)
\end{bmatrix}
= \begin{bmatrix}
\frac{-1}{M\iotabar} & \frac{1}{M} \\
-1 & 0
\end{bmatrix}
\begin{bmatrix}
\theta_B \\
\zeta_B
\end{bmatrix}
~&\text{for }M\neq0, N=0 \\
\begin{bmatrix}
\alpha \\
h(\rho,\alpha,\eta)
\end{bmatrix}
= \begin{bmatrix}
\frac{1}{N-M\iotabar} & \frac{-\iotabar}{N-M\iotabar} \\
-1 & \frac{N}{M}
\end{bmatrix}
\begin{bmatrix}
\theta_B \\
\zeta_B
\end{bmatrix}
~&\text{for }M\neq0, N\neq0
\end{align}
\end{subequations}
which can be inverted to yield
\begin{subequations}
\label{eq:mapping}
\begin{align}
\label{eq:map_poloidal}
\begin{bmatrix}
\theta_B \\
\zeta_B
\end{bmatrix}
= \begin{bmatrix}
N & \frac{\iotabar}{N} \\
0 & \frac{1}{N}
\end{bmatrix}
\begin{bmatrix}
\alpha \\
h(\rho,\alpha,\eta)
\end{bmatrix}
~&\text{for }M=0, N\neq0 \\
\label{eq:map_toroidal}
\begin{bmatrix}
\theta_B \\
\zeta_B
\end{bmatrix}
= \begin{bmatrix}
0 & -1 \\
M & \frac{-1}{\iotabar}
\end{bmatrix}
\begin{bmatrix}
\alpha \\
h(\rho,\alpha,\eta)
\end{bmatrix}
~&\text{for }M\neq0, N=0 \\
\label{eq:map_helical}
\begin{bmatrix}
\theta_B \\
\zeta_B
\end{bmatrix}
= \begin{bmatrix}
N & \frac{M\iotabar}{N-M\iotabar} \\
M & \frac{M}{N-M\iotabar}
\end{bmatrix}
\begin{bmatrix}
\alpha \\
h(\rho,\alpha,\eta)
\end{bmatrix}
~&\text{for }M\neq0, N\neq0.
\end{align}
\end{subequations}

The mapping parameters $x_{lmn}$ are used to compute $h(\rho,\eta,\alpha)$ through Eq. \ref{eq:h_comp}, and then Eq. \ref{eq:mapping} gives the corresponding Boozer coordinates such that $B_{\mathrm{eq}}(\rho,\eta,\alpha)$ can be evaluated in Boozer coordinates as $B_{\mathrm{eq}}(\theta_B,\zeta_B)$. 
These values of the equilibrium magnetic field strength are then used to compute the deviation from omnigenity in Eq. \ref{eq:f_om}. 
Note that Eq. \ref{eq:map_helical} is undefined when $\iotabar = \frac{N}{M}$, but this case is physically uninteresting because it corresponds to field lines being parallel to the $B_{\mathrm{max}}$ contour (for example, an axisymmetric vacuum field with no rotational transform). 
If $M=0$ then it must be the case that $N\neq0$, so Eq. \ref{eq:map_poloidal} is also always well defined. 
See figure \ref{fig:model} for an illustration of how an omnigenous magnetic field is constructed and mapped to Boozer coordinates. 

\subsection{Results}

It can be shown that this model is guaranteed to yield a perfectly omnigenous target magnetic field. 
Since the straight field line coordinate system has the property $\Delta\theta_B = \iotabar\Delta\zeta_B$ along a field line and $\iotabar$ is constant on each flux surface, and also because $h(\theta_B,\zeta_B)$ is always a linear combination of the Boozer angles, the bounce distance is directly proportional to the change in $h$: $\delta\propto\Delta h$. 
The bounce points on any field line are always $\pm\eta$ by construction, since Eq. \ref{eq:well} was chosen to be an even function of $\eta$. 
Therefore the bounce distance can be computed as: 
\begin{subequations}
\begin{align}
\delta \propto \Delta h &= h(\rho,+\eta,\alpha) - h(\rho,-\eta,\alpha) \\
&= 4\eta + \sum_{l=0}^{L_{\rho}} \sum_{m=0}^{M_{\eta}} \sum_{n=-N_{\alpha}}^{N_{\alpha}} x_{lmn} \Big[ \\
&T_{l}(2\rho-1) F_{n N_{\mathrm{FP}}}(\alpha) \Big( F_{m}(+\eta) - F_{m}(-\eta) \Big) \Big] \\
&= 4\eta.
\end{align}
\end{subequations}
In the last step, the Fourier series in $\eta$ cancel each other out because $\sum_{m\ge0} F_{m}(\eta)$ is an even function of $\eta$. 
This ensures that the bounce distances set by the parameters $B_{ij}$ will be preserved across all field lines, regardless of the variation introduced by the parameters $x_{lmn}$. 
However, there are bounds on the allowable amplitudes of these coefficients to ensure Eq. \ref{eq:h_comp} is a single-valued function. 

Since the basis functions span the full domain, this model can approximate any omnigenous magnetic field in the limit of increasing resolution (although this work is restricted to fields with a single magnetic well). 
Whether a particular omnigenous target can be well approximated by an equilibrium solution is an unresolved question. 
The total number of coefficients used to describe an omnigenous magnetic field in Eq. \ref{eq:h_comp} is $(L_\rho+1) (M_\eta+1) (2 N_\alpha + 1)$. 
Quasi-symmetry corresponds to the condition $N_\alpha=0$, so that the magnetic well shape is identical on all field lines and $B = B(h(\rho,\eta))$. 
Assuming $L_\rho = M_\eta = N_\alpha$, the dimension of the full omnigenous parameter space scales as $\mathcal{O}(M_\eta^3)$ while the quasi-symmetric subspace has a dimension of $\mathcal{O}(M_\eta^2)$. 
This scaling reveals that quasi-symmetry is only a small set of the possible solutions with good particle confinement, and motivates the discovery of stellarator equilibria in the much broader omnigenous design space. 
[This analysis is independent of the dimension of the magnetic well parameterization $B_{ij}$.] 
Stellarator symmetry imposes no constraints on the parameters $x_{lmn}$ due to the nonlinear nature of the coordinate mapping in Eq. \ref{eq:h}. 
This implies that the size of the parameter space is not reduced by stellarator symmetry, but consequently no additional computational complexity is required if it is not assumed. 

\section{Examples}
\label{sec:examples}

Numerical examples from each class of omnigenity are presented: the three types of quasi-symmetry (QP, QH, and QA) and their more general omnigenous counterparts (OP, OH, and OT, respectfully). 
These six solutions were obtained in DESC using the following optimization procedure. 
First, an initial guess was generated from the near-axis expansion code pyQIC \citep{pyQIC} (for the OP and QP case) or pyQSC \citep{Landreman2019a,Landreman2019b} (for the other cases). 
It was evaluated at an aspect ratio of $R_0/a \approx 20$, since the near-axis expansions are most accurate at high aspect ratios. 
This configuration was re-solved in DESC while preserving the $\mathcal{O}(\rho)$ behavior of the near-axis expansion solution. 
The equilibrium was then optimized using a multi-objective cost function of the form
\begin{equation}
\label{eq:cost}
\mathbf{f} = \begin{bmatrix}
f_{\mathrm{om}}(\rho,\eta,\alpha) \\
\sigma f_{\mathrm{eq}}(\rho,\theta,\zeta)
\end{bmatrix}
\end{equation}
where $f_{\mathrm{om}}(\rho,\eta,\alpha)$ are the omnigenity errors from Eq. \ref{eq:f_om}, $f_{\mathrm{eq}}(\rho,\theta,\zeta)$ are the equilibrium force balance errors as described in \cite{Dudt2020}, and $\sigma$ is a relative weighting between the two objective functions. 
This penalty method was used instead of the traditional optimization approach of re-solving the fixed-boundary equilibrium at each iteration \citep{Dudt2023} because it can better handle the near-axis constraints, and is equivalent in the limit $\sigma\to\infty$. 
Omnigenity was targeted on five flux surfaces $\rho = 0.2,~0.4,~0.6,~0.8,~1.0$ and the shape of the magnetic well $B(\eta)$ on each surface was parameterized by a monotonic spline with eight knots. 
The force balance errors were evaluated throughout the plasma volume. 
The residuals in Eq. \ref{eq:cost} were minimized with a least-squares trust region algorithm. 
The total magnetic flux, equilibrium profiles, and $\mathcal{O}(\rho)$ behavior from the near-axis expansion solution were constrained during the optimization. 
This process was repeated three times, with the numerical resolutions and $\sigma$ increasing at each iteration. 
Finally, the optimal configuration underwent a fixed-boundary solve to ensure its volume-averaged normalized force balance error was $<1\%$ \citep{Panici2023}. 

\begin{figure}
\centering
\includegraphics[width=\textwidth]{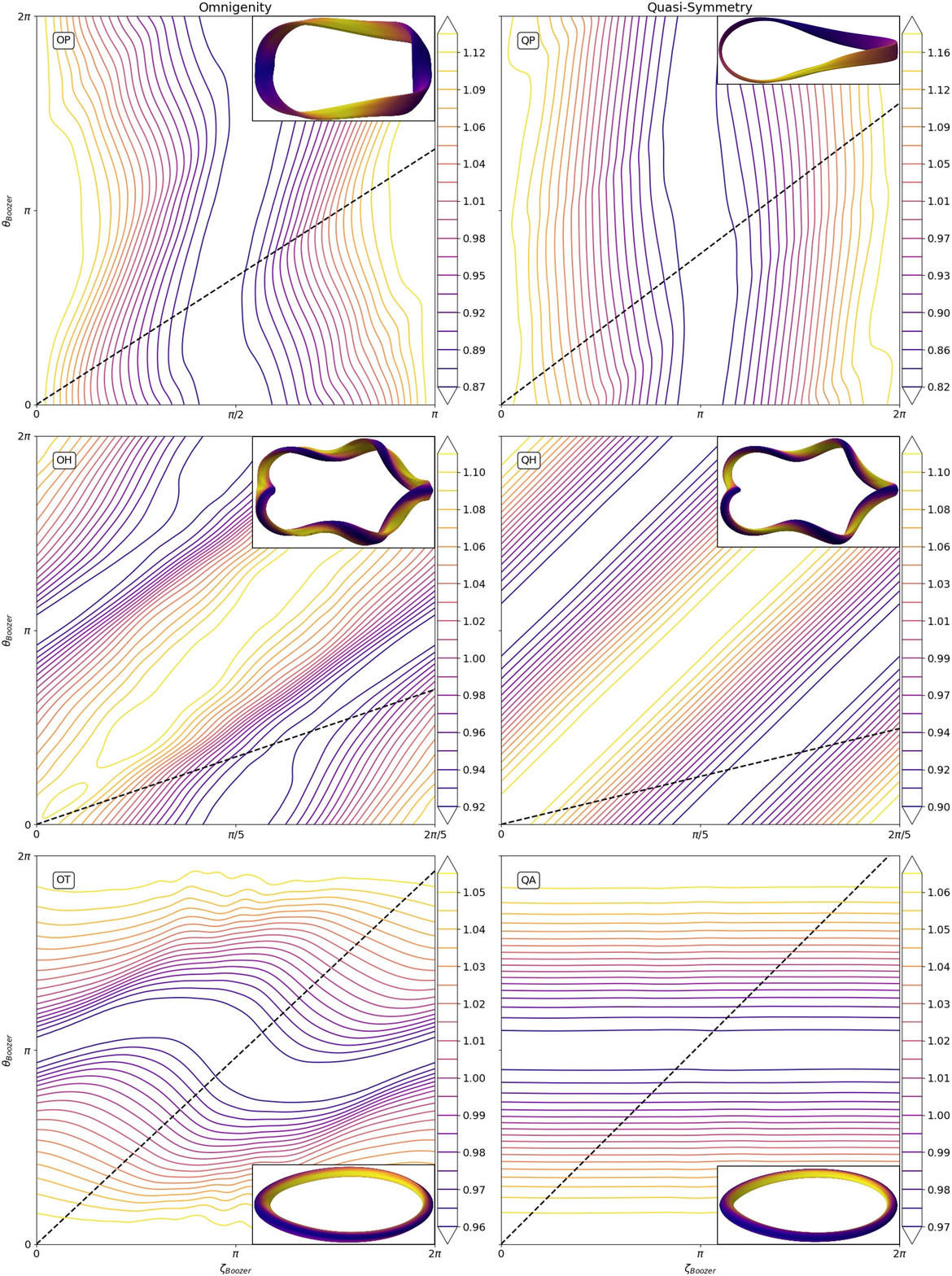}
\caption{Contours of constant magnetic field magnitude in Boozer coordinates at the boundary surface $\rho=1$ for a single field period. 
The magnetic field line at $\alpha=0$ is shown as a black dashed line in each plot.
The insets show a three-dimensional view of the field strength on each surface.}
\label{fig:fields}
\end{figure}

Each optimization took less than thirty minutes to run on a single NVIDIA A100 GPU. 
All of the examples have a major radius of $1~\mathrm{m}$ and an average magnetic field strength on axis of $1~\mathrm{T}$ by construction. 
The results are stellarator symmetric, but this was a design choice and not a limitation of the method. 
Figure \ref{fig:fields} displays the magnetic field strength in Boozer coordinates on the $\rho=1$ surface for these equilibria. 
These plots clearly reveal the desired helicity of the (straight) $B_{\mathrm{max}}$ contours for each case, and they help picture how the bounce distances are roughly equal on every field line (satisfying the condition $\frac{\partial\delta}{\partial\alpha} = 0$). 
A scalar omnigenity error is defined as the mean of the local relative errors:
\begin{equation}
\frac{1}{K} \sum_{i=1}^K \frac{\big| B_{\mathrm{eq}}(\rho_i,\eta_i,\alpha_i) - B(\rho_i,\eta_i) \big|}{B(\rho_i,\eta_i)}
\end{equation}
and the normalized quasi-symmetry error in Boozer coordinates is defined as \citep{Rodriguez2022}: 
\begin{equation}
\frac{\sqrt{\sum_{\frac{n}{m}\neq\frac{N}{M}} B^2_{mn}}}{\sqrt{\sum_{m,n}B^2_{mn}}}.
\end{equation}
These errors along with the aspect ratio and maximum elongation of each solution are provided in table \ref{tab:configs}. 
The aspect ratio was indirectly constrained during the optimization by fixing the magnetic axis shape, magnetic field strength on axis, and total magnetic flux; the elongation was not constrained. 

\begin{figure}
\centering
\includegraphics[width=0.5\textwidth]{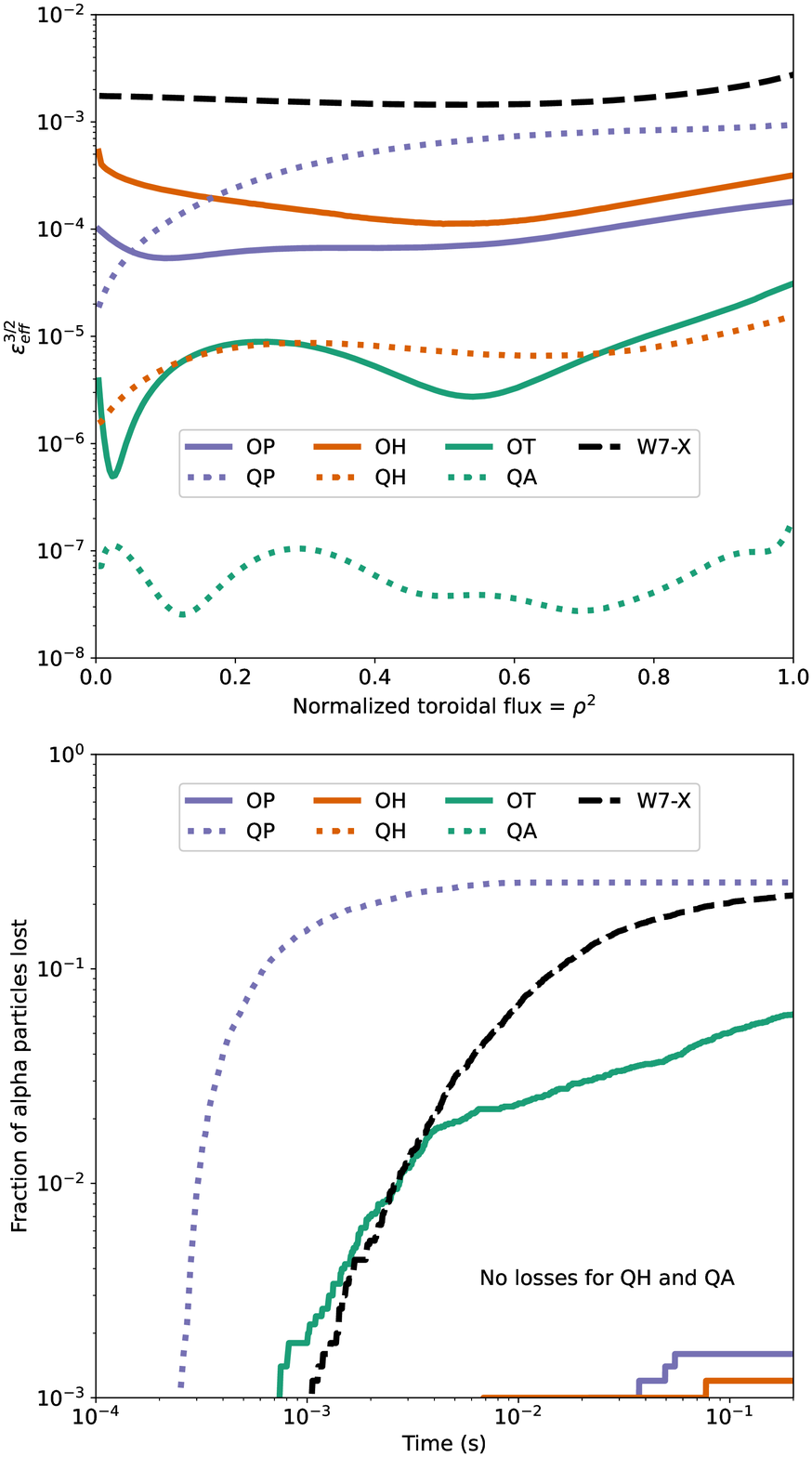}
\caption{Measures of confinement. 
The quasi-symmetric solutions are plotted as dotted lines, while the more general omnigenous solutions are plotted as solid lines. 
Solutions with the same type of helicity are plotted in the same color. 
The Wendelstein 7-X configuration with $\beta=4\%$ is plotted in a black dashed line for comparison \citep{Landreman}. 
Top: Neoclassical collisional transport magnitude, computed by NEO. 
Bottom: Collisionless losses of fusion-born alpha particles initialized at $\rho=0.5$, computed by SIMPLE. 
All configurations were scaled to the minor radius and magnetic field strength of ARIES-CS.}
\label{fig:confinement}
\end{figure}

As another measure of the levels of omnigenity achieved, the effective ripple $\epsilon_{\mathrm{eff}}^{3/2}$ -- the magnitude of the neoclassical collisional transport in the $\frac{1}{\nu}$ regime, where $\nu$ is the collision frequency -- is computed by NEO \citep{Nemov1999} for each solution. 
As another measure of confinement, the time history of fusion-born alpha particles lost in a reactor scale device is also computed by SIMPLE \citep{Albert2020}. 
In a perfectly omnigenous magnetic field, $\epsilon_{\mathrm{eff}}^{3/2}$ should be zero and no particles should be lost besides those with banana orbits that are wide enough to leave the plasma volume. 
Both of these confinement measures are plotted in figure \ref{fig:confinement}. 
The numerical examples shown here were only optimized for omnigenity and equilibrium quality, and other factors that may be important for a fusion reactor design were not considered. 
More details about each optimized solution are provided in the following subsections. 

\begin{table}
\caption{Comparison between the six omnigenous solutions of geometric parameters and magnetic field errors at the $\rho=1$ surface. 
The aspect ratios are calculated using the same formula as the VMEC equilibrium code \citep{Hirshman1983}. 
The omnigenity errors were computed with different collocation points than those used in the optimization.}
\label{tab:configs}
\begin{center}
\def~{\hphantom{0}}
\begin{tabular}{ccccccc}
\hline
 & Aspect & Maximum & Quasi-symmetry & Omnigenity \\
 & Ratio & Elongation & Error & Error \\
\hline
OP & 19.35 & 6.70 & 3.73E-2 & 8.96E-4 \\
QP & 17.91 & 10.10 & 1.78E-2 & 1.02E-2 \\
OH & 18.27 & 5.74 & 2.24E-2 & 3.15E-3 \\
QH & 18.24 & 2.48 & 1.62E-3 & 8.64E-4 \\
OT & 19.56 & 3.37 & 1.41E-2 & 2.74E-3 \\
QA & 19.44 & 1.95 & 2.81E-4 & 1.78E-4
\end{tabular}
\end{center}
\end{table}

\subsection{Poloidal Contours}

Omnigenity with poloidally closed $B$ contours is characterized by a helicity of $M=0$. 
These examples of OP and QP are vacuum equilibria ($\langle\beta\rangle=0$) that were based on initial guesses generated by pyQIC. 
The OP case is a two field period configuration with a mirror ratio of $\Delta=0.13$ and the QP case has a single field period with $\Delta=0.17$, where the mirror ratio is defined as
\begin{equation}
\Delta = \frac{B_\mathrm{max}-B_\mathrm{min}}{B_\mathrm{max}+B_\mathrm{min}}
\end{equation}
on the magnetic axis. 
In addition to the flux surface geometry, the omnigenity parameters $B_{ij}$ and $x_{lmn}$ were also considered free variables during the optimization. 
A resolution of $L_\rho = M_\eta = N_\alpha = 4$ was used for the OP case, while the QP case used a resolution of $L_\rho = M_\eta = 4$, $N_\alpha = 0$ to enforce quasi-symmetry. 
The OP solution has worse QS error than the QP example, as expected, but achieves an order of magnitude lower omnigenity error and effective ripple, and two orders of magnitude lower fast particle losses (see table \ref{tab:configs} and figure \ref{fig:confinement}). 

\subsection{Helical Contours}

The typical case of omnigenity with helically closed $B$ contours is $M=1$ and $N=N_{\mathrm{FP}}$. 
These examples of OH and QH are five field period vacuum equilibria ($\langle\beta\rangle=0$) using the same initial guess from pyQSC. 
As in the previous poloidal examples, the parameters $B_{ij}$ describing the magnetic well shapes on each surface were considered free optimization variables. 
However, in these runs the coordinate mapping of Eq. \ref{eq:h_comp} was prescribed: $x_{lmn}=0 ~\forall~ l,m,n$ in the QH case, and $x_{0,1,+1}=-\frac{\pi}{6}$ was the only non-zero parameter in the OH case. 
This parameter value for the OH example was chosen such that the resulting field, shown in the middle left panel of figure \ref{fig:fields}, qualitatively resembles the $M=1$, $N=4$ omnigenous field example constructed by \cite{Landreman2012}. 
In both cases, all of the $x_{lmn}$ variables were held fixed during the optimizations to ensure that they converged to a quasi-symmetric and general omnigenous magnetic field, respectively. 
The QS error of the OH solution is an order of magnitude higher than that of the QH result, verifying that it is far from quasi-symmetry (see table \ref{tab:configs}). 

\subsection{Toroidal Contours}

$N=0$ describes omnigenity with toroidally closed $B$ contours. 
As explained by \cite{Landreman2012}, ``it is difficult to construct $N=0$ omnigenous fields that depart strongly from quasi-symmetry \dots because the $B$ contours in a $N=0$ quasi-symmetric field are already nearly parallel to the field lines when $\iotabar/N_{\mathrm{FP}}<1$''. 
These examples of OT and QA are single field period equilibria based on the same initial guess from pyQSC, with a finite toroidal current profile and a volume averaged normalized plasma pressure of $\langle\beta\rangle=1.2\%$. 
The plasma current provides additional rotational transform, which allows for an OT case that is far from quasi-symmetry (as seen in the top right panel of figure \ref{fig:fields} and by the relatively high QS error in table \ref{tab:configs}). 
These toroidal solutions were found using the same optimization process as the previous helical examples, except that the coordinate mapping for the target OT field had $x_{0,1,-1}=\frac{\pi}{6}$ as the only non-zero parameter. 
As in the helical examples, all of the $x_{lmn}$ variables were constrained to their target values while all of the $B_{ij}$ were free variables during the optimization. 

\section{Discussion}
\label{sec:discussion}

This is the first demonstration of equilibria solutions that closely approximate every class of omnigenous magnetic fields. 
All of the $B$ contour plots in figure \ref{fig:fields} have their expected helicities, and it can be seen that the distances between equal contours are roughly equivalent along any field line. 
The average relative omnigenity errors listed in table \ref{tab:configs} are all $\lessapprox1\%$, indicating the optimized equilibria closely resemble their omnigenous targets. 
The good confinement properties of these omnigenous solutions at moderate aspect ratio are verified by the neoclassical collisional transport magnitudes $\epsilon_{\mathrm{eff}}^{3/2}$ and the loss fractions displayed in figure \ref{fig:confinement}. 
These relatively low values of both confinement metrics are all comparable in magnitude to previous optimized configurations \citep{Landreman2022a}. 
In particular, the OT solution has very low effective ripple and the OP and OH solutions have almost no alpha particle losses -- on par with the QS results. 
This indicates the existence of stellarator equilibria other than OP that can approach omnigenity without being quasi-symmetric. 

The OH and OT results do have higher omnigenity errors and worse confinement than their respective quasi-symmetric counterparts, but the target parameters for these examples were chosen arbitrarily and other omnigenous fields with better confinement properties might exist. 
Omnigenous fields that are not quasi-symmetric are impossible to achieve exactly, and more examples are needed to understand if they can generally be approximated well. 
The OT result also has surprisingly high alpha particle losses considering its very low effective ripple, exemplifying the complicated relationship between these two confinement metrics. 
This could be explained by wide banana orbits that extend beyond the plasma volume, but more investigation is required to determine the cause. 
The outlier of all the results is the QP solution, which cannot exist near the magnetic axis \citep{Plunk2019} and is known to be difficult to achieve far from the axis without highly elongated surfaces. 
This QP example may have unrealistic geometry for a reactor, but it is still encouraging to find a quasi-poloidally symmetric configuration with similar levels of confinement as the Wendelstein 7-X experiment. 

The purpose of the examples in Section \ref{sec:examples} are to establish solutions that are representative of each type of omnigenity that can exist. 
The model presented in Section \ref{sec:method} and the optimization capabilities of DESC provide flexibility for a variety of applications. 
The parameters $B_{ij}$ and $x_{lmn}$ can either be constrained to target a specific location in the omnigenity design space (as in the OH, OT, QH, and QA examples), or they can be free optimization variables (as in the OP and QP examples). 
Providing more degrees of freedom would likely result in the discovery of configurations with even better confinement properties or that satisfy additional optimization objectives. 
Eliminating the assumption of stellarator symmetry is a logical step in this direction, as it doubles the size of the omnigenity solution space that can be explored. 
As another example, the boundary condition of Eq. \ref{eq:bc} could be relaxed and only satisfied approximately; this would violate the already impossible goal of exact omnigenity, but might improve the solution quality overall. 
Alternatively, constraining most of the parameters would enable a systematic scan of the full omnigenity design space. 
Mapping this space, like the surveys of quasi-symmetry made in \cite{Rodriguez2023,Landreman2022b}, is an exciting direction for future research. 
Other ideas for future work include investigating how the larger omnigenity phase space (in contrast to QS) could simplify stellarator coil geometry and yield more promising candidates for fusion power plants. 

The code and data that support the findings of this study are openly available in the DESC repository at \url{https://github.com/PlasmaControl/DESC}. 
José Luis Velasco and collaborators have presented a similar method to parameterize a subset of omnigenous magnetic fields with poloidal contours, and they are thanked for their contributions. 
This work was supported by the U.S. Department of Energy under contract numbers DE-AC02-09CH11466,
DE-SC0022005, and Field Work Proposal No. 1019. 
The United States Government retains a non-exclusive, paid-up, irrevocable, world-wide license to publish or reproduce the published form of this manuscript, or allow others to do so, for United States Government purposes. 
The authors report no conflicts of interest. 

\bibliographystyle{jpp}
\bibliography{bibliography}

\end{document}